%% file: Google_Matrices.tex
\documentclass[12pt,a4paper]{article}
\usepackage{amsmath}
\usepackage{latexsym}
\makeatletter
\def\rddots{\mathinner{\mkern1mu\raise\p@%
    \vbox{\kern7\p@\hbox{.}}\mkern2mu%
    \raise4\p@\hbox{.}\mkern2mu\raise7\p@\hbox{.}\mkern1mu}}
\makeatother
\makeatletter
   
   \@addtoreset{equation}{section}
\makeatother
\begin{document}

\title{\sl A Superintroduction to Google Matrices for Undergraduates}
\author{
  Kazuyuki FUJII
  \thanks{E-mail address : fujii@yokohama-cu.ac.jp }\quad and\ \ 
  Hiroshi OIKE
  \thanks{E-mail address : oike@tea.ocn.ne.jp}\\
  ${}^{*}$International College of Arts and Sciences\\
  Yokohama City University\\
  Yokohama, 236--0027\\
  Japan\\
  ${}^{\dagger}$Takado\ 85--5,\ Yamagata, 990--2464\\
  Japan\\
  }
\date{}
\maketitle
\begin{abstract}
  In this paper we consider so-called Google matrices and show 
that all eigenvalues ($\lambda$) of them have a fundamental  
property $|\lambda|\leq 1$. The stochastic eigenvector 
corresponding to $\lambda=1$ called the PageRank vector 
plays a central role in the Google's software. 
We study it in detail and present some important problems.
 
  The purpose of the paper is to make {\bf the heart of Google}  
clearer for undergraduates.
\end{abstract}

\vspace{5mm}\noindent

\vspace{3mm}\noindent

\section{Introduction}
Google is one of important tools to analyze Modern Society. 
In this paper we want to explain a secret of Google, 
which is ``the heart of Google's software", to undergraduates. 

Although we are not experts of IT (Information Technology) 
the secret is clearly expressed in terms of Linear Algebra 
in Mathematics. However, it is almost impossible to solve 
the linear algebra version explicitly, so we need some 
approximate method.

First, we give a fundamental lemma to understand a Google 
matrix (see the definition in the text) and present an 
important problem to define a realistic Google matrix 
(in our terminology). The problem is a challenging one 
for young researchers. For such a matrix we can use 
the power method to obtain the PageRank vector.

Second, we pick up an interesting example in \cite{DA} and 
calculate it thoroughly by use of MATHEMATICA. 
A good example and a thorough calculation 
help undergraduates to understand. 

Last, we show an example which does not give the 
PageRank vector in terms of the power method with 
usual initial vector when $H$ is not a realistic Google matrix. 
For this case we treat the power method with another 
initial vector and present a general problem.

We expect that undergraduates will cry out ``I got Google !" 
after reading the paper.

\section{Main Result}
We introduce a Google matrix (realistic Google matrix) and 
study its key property. 

We consider a collection of web pages 
with links (for example, a homepage and some 
homepages cited in it). See the figure in the next 
section (eight web pages with several links). 

If a page has $k$ links we give the equal weight $\frac{1}{k}$ 
to each link and construct a column vector consisting of 
these weights. 
See the figure once more. For example, 
since the page $4$ links to the pages $2$, $5$ and $6$ (three 
links) each weight is $\frac{1}{3}$. 
Therefore we obtain the column vector like

\[
\mbox{page\ 4}\ \longrightarrow \ 
\left(
\begin{array}{c}
0 \\
\frac{1}{3} \\
0 \\
0 \\
\frac{1}{3} \\
\frac{1}{3} \\
0 \\
0
\end{array}
\right)
\hspace{-3mm}
\left.
\begin{array}{l}
{}\\
<2 \\
{} \\
{} \\
<5 \\
<6 \\
{} \\
{}
\end{array}
\right.
\]
As a result, the collection of web pages gives a square matrix
\begin{equation}
H=(H_{ij});
\quad H_{ij}\geq 0,
\quad 
\sum_{i}H_{ij}=1
\end{equation}
which we will call a Google matrix. Note that $H_{ii}=0$ for 
all $i$ (we prohibit the self-citation). 
From the definition it is a {\bf sparse} matrix because the 
number of links starting from a webpage is in general small 
compared to the number of webpages.

If we set
\[
J=(1,1,\cdots,1)^{T}
\]
where $T$ is the transpose (of a vector or a matrix) then 
it is easy to see
\begin{equation}
H^{T}J=J
\end{equation}
because row vectors of $H^{T}$ are the transpose of 
column vectors of $H$ like
\[
\mbox{page\ 4} \longrightarrow 
\left(0,\frac{1}{3},0,0,\frac{1}{3},\frac{1}{3},0,0\right).
\]
From this we know that $1$ is an eigenvalue of $H^{T}$. 
By the way, the eigenvalues of $H$ are equal to those of 
$H^{T}$ because
\[
0=|\lambda E-H|=|\lambda E-H^{T}|
\]
, so we conclude that $1$ is just an eigenvalue of $H$.

Therefore, we have the equation
\begin{equation}
\label{eq:characteristic eq}
HI=I
\end{equation}
where we assume that the eigenvector $I$ is stochastic 
(the sum of all entries is $1$). 
This $I$ is called the PageRank vector and plays a central  
role in Google.

Now, we give a fundamental lemma to Google matrices :

\vspace{3mm}\noindent
{\bf Lemma}\ \ Let $\lambda$ be any eigenvalue of a
Google matrix $H$. Then we have
\begin{equation}
|\lambda|\leq 1.
\end{equation}

The proof is easy and is derived from the Gerschgorin's 
(circle) theorem \cite{SG}. Note that the eigenvalues of $H$ 
are equal to those of $H^{T}$ and the sum of all entries of 
each row is $1$ (see for example (\ref{eq:H^{T}})). 
Namely, 
\begin{equation}
\label{eq:entry}
\sum_{j=1}^{n}(H^{T})_{ij}=\sum_{j=1}^{n}H_{ji}=1
\quad \mbox{and}\quad (H^{T})_{ii}=H_{ii}=0
\end{equation}
for all $i$ and $j$.

We are in a position to state the Gerschgorin's theorem. 
Let $A=(a_{ij})$ be a $n\times n$ complex (real in our case) 
matrix, and we set
\[
R_{i}=\sum_{j=1,\ j\ne i}^{n}|a_{ij}|
\]
and 
\[
D(a_{ii}; R_{i})=\{z\in {\bf C}\ |\ |z-a_{ii}|\leq R_{i}\}
\]
for each $i$. 
This is a closed disc centered at $a_{ii}$ with radius $R_{i}$ 
called the Gerschgorin's disc.

\vspace{3mm}\noindent
{\bf Theorem (Gerschgorin)}\ \ For any eigenvalue $\lambda$ 
of $A$ we have
\begin{equation}
\lambda \in \bigcup_{i=1}^{n}D(a_{ii}; R_{i}).
\end{equation}

The proof is simple. Let us consider the equation
\begin{equation}
\label{eq:characteristic}
A{\bf x}=\lambda{\bf x}\quad ({\bf x}\ne {\bf 0})
\end{equation}
and $|x_{i}|$ be the maximum
\[
|x_{i}|=\mbox{max}\{|x_{1}|,|x_{2}|,\cdots,|x_{n}|\}\ne 0
\ \Longrightarrow \ 
|\frac{x_{j}}{x_{i}}|\leq 1.
\]
From (\ref{eq:characteristic}) we have
\[
\sum_{j=1}^{n}a_{ij}x_{j}=\lambda x_{i}
\Longleftrightarrow 
\sum_{j=1,\ j\ne i}^{n}a_{ij}x_{j}
=\lambda x_{i}-a_{ii}x_{i}
=(\lambda -a_{ii})x_{i}.
\]
$x_{i}\ne 0$ gives
\[
\lambda -a_{ii}=\sum_{j=1,\ j\ne i}^{n}a_{ij}\frac{x_{j}}{x_{i}}
\]
and we have
\[
|\lambda -a_{ii}|=|\sum_{j=1,\ j\ne i}^{n}a_{ij}\frac{x_{j}}{x_{i}}|
\leq \sum_{j=1,\ j\ne i}^{n}|a_{ij}\frac{x_{j}}{x_{i}}|
= \sum_{j=1,\ j\ne i}^{n}|a_{ij}||\frac{x_{j}}{x_{i}}|
\leq \sum_{j=1,\ j\ne i}^{n}|a_{ij}|
=R_{i}.
\]
This means $\lambda \in D(a_{ii}; R_{i})$ for some $i$ 
and completes the proof.

\vspace{3mm}
Finally, let us complete our lemma. In our case $H_{ij}\geq 0$, 
$H_{ii}=0$ and $R_{i}=1$ for all $i$ and $j$, so these give the result
\[
|\lambda|\leq 1
\]
for any eigenvalue $\lambda$ of $H$. This is indeed a fundamental 
property of Google matrices.

\vspace{3mm}
A comment is in order. The lemma must have been known. However, 
we could not find such a reference within our efforts.

Let us go ahead. In order to construct the eigenvector $I$  
in (\ref{eq:characteristic eq}) a method called {\bf the power method}  
is very convenient for a sparse matrix $H$ of huge size. 
To calculate the characteristic polynomial is actually impossible.

The method is very simple, \cite{DA}. A sequence $\{I_{n}\}$ is defined 
recurrently by
\begin{equation}
I_{n}=HI_{n-1}\quad \mbox{and}\quad I_{0}={\bf e_{1}}
\end{equation}
where the initial vector is ${\bf e_{1}}=(1,0,\cdots,0)^{T}$, which is 
usually standard. This is also rewritten as
\[
I_{n}=H^{n}I_{0}=H^{n}{\bf e_{1}}.
\]
If $\{I_{n}\}$ converges to $I$ then we obtain the equation 
(\ref{eq:characteristic eq}) like
\[
HI
=H(\lim_{n\rightarrow \infty}I_{n})
=\lim_{n\rightarrow \infty}HI_{n}
=\lim_{n\rightarrow \infty}I_{n+1}
=I.
\]

In order that the power method works correctly some 
assumption on $H$ is required. Namely, 

\vspace{3mm}\noindent
($\heartsuit$)\ \ For a set of eigenvalues 
$\{\lambda_{1}=1,\lambda_{2}, \cdots,\lambda_{n}\}$ we assume
\begin{equation}
\label{eq:realistic}
\lambda_{1}=1>|\lambda_{2}|\geq \cdots \geq |\lambda_{n}|.
\end{equation}
Note that $1$ is a simple root. The assumption may be strong. 

If a Google matrix $H$ satisfies (\ref{eq:realistic}) we 
call $H$ a {\bf realistic} Google matrix. 
Now, let us present an important

\vspace{3mm}\noindent
{\bf Problem}\ \ For a huge sparse matrix $H$ propose a method 
to find or to estimate the second eigenvalue $\lambda_{2}$ 
without calculating the characteristic polynomial. 

\vspace{3mm}
As far as we know such a method has not been given in 
Mathematical Physics or in Quantum Mechanics. 
This is a challenging problem for mathematical physicists.

\section{Example}
We consider an interesting example given in \cite{DA} and 
calculate it thoroughly by use of MATHEMATICA. 
A good example helps undergraduates to understand 
a model deeply.

In this section we need some results from Linear Algebra, 
so see for example \cite{IS} or \cite{KF} (we don't know 
a standard textbook of Linear Algebra in Europe or 
America or etc).

\noindent
Example : a collection of web pages with links\footnote{
It is not easy for us to draw a (free) curve by use of 
the free soft WinTpic.}

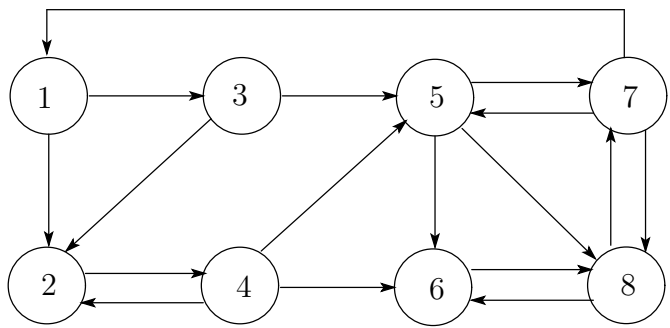
\vspace{3mm}
\begin{center}
\input{google-figure-1.tex}
\end{center}

\vspace{3mm}\noindent
The Google matrix for this graph is given by
\begin{equation}
\label{eq:H}
H=
\left(
\begin{array}{cccccccc}
0 & 0 & 0 & 0 & 0 & 0 & \frac{1}{3} & 0  \\
\frac{1}{2} & 0 & \frac{1}{2} & \frac{1}{3} & 0 & 0 & 0 & 0  \\
\frac{1}{2} & 0 & 0 & 0 & 0 & 0 & 0 & 0  \\
0 & 1 & 0 & 0 & 0 & 0 & 0 & 0  \\
0 & 0 & \frac{1}{2} & \frac{1}{3} & 0 & 0 & \frac{1}{3} & 0  \\
0 & 0 & 0 & \frac{1}{3} & \frac{1}{3} & 0 & 0 & \frac{1}{2}  \\
0 & 0 & 0 & 0 & \frac{1}{3} & 0 & 0 & \frac{1}{2}  \\
0 & 0 & 0 & 0 & \frac{1}{3} & 1 & \frac{1}{3} & 0
\end{array}
\right)
\end{equation}
and its transpose is
\begin{equation}
\label{eq:H^{T}}
H^{T}=
\left(
\begin{array}{cccccccc}
0 & \frac{1}{2} & \frac{1}{2} & 0 & 0 & 0 & 0 & 0 \\
0 & 0 & 0 & 1 & 0 & 0 & 0 & 0  \\
0 & \frac{1}{2} & 0 & 0 & \frac{1}{2} & 0 & 0 & 0 \\
0 & \frac{1}{3} & 0 & 0 & \frac{1}{3} & \frac{1}{3} & 0 & 0 \\
0 & 0 & 0 & 0 & 0 & \frac{1}{3} & \frac{1}{3} & \frac{1}{3}  \\
0 & 0 & 0 & 0 & 0 & 0 & 0 & 1 \\
\frac{1}{3} & 0 & 0 & 0 & \frac{1}{3} & 0 & 0 & \frac{1}{3} \\
0 & 0 & 0 & 0 & 0 & \frac{1}{2} & \frac{1}{2} & 0
\end{array}
\right).
\end{equation}

If we define a stochastic vector
\[
J=\left(\frac{1}{8},\frac{1}{8},\frac{1}{8},\frac{1}{8},\frac{1}{8},\frac{1}{8},\frac{1}{8},\frac{1}{8}\right)^{T}
\]
it is easy to see
\begin{equation}
H^{T}J=J.
\end{equation}

Let us study $H$ from the mathematical view point 
by use of MATHEMATICA.  The characteristic polynomial 
of $H$ is given by

\begin{eqnarray}
f(\lambda)
&=&|\lambda E-H| \nonumber \\
&=&
\left|
\begin{array}{cccccccc}
\lambda & 0 & 0 & 0 & 0 & 0 & -\frac{1}{3} & 0  \\
-\frac{1}{2} & \lambda & -\frac{1}{2} & -\frac{1}{3} & 0 & 0 & 0 & 0  \\
-\frac{1}{2} & 0 & \lambda & 0 & 0 & 0 & 0 & 0  \\
0 & -1 & 0 & \lambda & 0 & 0 & 0 & 0  \\
0 & 0 & -\frac{1}{2} & -\frac{1}{3} & \lambda & 0 & -\frac{1}{3} & 0  \\
0 & 0 & 0 & -\frac{1}{3} & -\frac{1}{3} & \lambda & 0 & -\frac{1}{2}  \\
0 & 0 & 0 & 0 & -\frac{1}{3} & 0 & \lambda & -\frac{1}{2}  \\
0 & 0 & 0 & 0 & -\frac{1}{3} & -1 & -\frac{1}{3} & \lambda
\end{array}
\right| \nonumber \\
&=&\lambda(\lambda -1)\left(\lambda^{6}+\lambda^{5}-
\frac{1}{9}\lambda^{4}-\frac{1}{6}\lambda^{3}+\frac{7}{108}\lambda^{2}+
\frac{11}{216}\lambda +\frac{1}{72}\right).
\end{eqnarray}
The exact solutions are $\{\lambda_{1}=1, \lambda_{8}=0 \}$ and 
approximate ones (we round off a real number to five decimal 
places like $-0.87021\cdots\ =-0.8702$) are given by
\begin{eqnarray*}
\lambda_{2}&=&-0.8702,\ \ 
\lambda_{3}=-0.5568, \\
\lambda_{4}&=&0.4251-0.2914i, \ \ 
\lambda_{5}=0.4251+0.2914i, \\
\lambda_{6}&=&-0.2116-0.2512i, \ \ 
\lambda_{7}=-0.2116+0.2512i.
\end{eqnarray*}
From these we have
\begin{equation}
\lambda_{1}=1>|\lambda_{2}|>|\lambda_{3}|>|\lambda_{4}|=|\lambda_{5}|>
|\lambda_{6}|=|\lambda_{7}|>\lambda_{8}=0.
\end{equation}
$H$ becomes a realistic Google matrix from (\ref{eq:realistic}).

Moreover, the eigenvector for $\lambda_{1}=1$ is given by
\[
\hat{I}=(24, 27, 12, 27, 39, 81, 72, 118)^{T}.
\]
To check this (by hand) is not difficult and good exercise 
for undergraduates. 
Since the sum of all entries of $\hat{I}$ is 400 
the stochastic eigenvector (= the PageRank vector) 
$I$ becomes

\begin{equation}
I=
\left(
\begin{array}{c}
\frac{24}{400}  \\
\frac{27}{400}  \\
\frac{12}{400}  \\
\frac{27}{400}  \\
\frac{39}{400}  \\
\frac{81}{400}  \\
\frac{72}{400}  \\
\frac{118}{400}
\end{array}
\right)
=
\left(
\begin{array}{l}
0.06 \\
0.0675 \\
0.03 \\
0.0675 \\
0.0975 \\
0.2025 \\
0.18 \\
0.295
\end{array}
\right).
\end{equation}

\vspace{3mm}\noindent
As a result, the ranking of webpages becomes
\begin{equation}
\mbox{page 8}>\mbox{page 6}>\mbox{page 7}>\mbox{page 5}>\mbox{page 2}=\mbox{page 4}>\mbox{page 1}>\mbox{page 3}.
\end{equation}
See the figure once more.

\vspace{3mm}
Here, let us show the power method to obtain the 
PageRank vector $I$, which is very useful if a realistic 
Google matrix is huge. 
A sequence $\{I_{n}\}$ is defined as
\[
I_{n}=HI_{n-1}\quad \mbox{and}\quad I_{0}=(1,0,0,0,0,0,0,0)^{T}
\]
or
\[
I_{n}=H^{n}I_{0}.
\]

If the condition $|\lambda_{2}|<1$ holds then we have 
\[
\lim_{n\rightarrow \infty}I_{n}=I
\]
because $H$ can be diagonalized to be
\[
H=S\mbox{diag}(1,\lambda_{2},\cdots,\lambda_{8})S^{-1}
\Longrightarrow 
H^{n}=S\mbox{diag}(1,\lambda_{2}^{n},\cdots,\lambda_{8}^{n})S^{-1}
\]
with a matrix $S$ consisting of eigenvectors. The speed of 
convergence depends on $|\lambda_{2}|$.

Let us list the calculation (rule : a real number is 
rounded off to five decimal places) :
\begin{equation}
I_{40}=
\left(
\begin{array}{c}
0.0601 \\
0.0675 \\
0.0299 \\
0.0676 \\
0.0976 \\
0.2022 \\
0.1797 \\
0.2954
\end{array}
\right),
\
I_{45}=
\left(
\begin{array}{c}
0.0600 \\
0.0675 \\
0.0300 \\
0.0675 \\
0.0975 \\
0.2024 \\
0.1800 \\
0.2951
\end{array}
\right),
\
I_{50}=
\left(
\begin{array}{c}
0.0600 \\
0.0675 \\
0.0300 \\
0.0675 \\
0.0975 \\
0.2024 \\
0.1799 \\
0.2951
\end{array}
\right),
\
I_{55}=
\left(
\begin{array}{c}
0.0600 \\
0.0675 \\
0.0300 \\
0.0675 \\
0.0975 \\
0.2025 \\
0.1800 \\
0.2950
\end{array}
\right)\equiv I.
\end{equation}

\vspace{3mm}\noindent
The result must be related to the powers of $|\lambda_{2}|=0.87$ 
like
\begin{equation}
(0.87)^{40}=0.0038,
\quad
(0.87)^{45}=0.0019,
\quad
(0.87)^{50}=0.0009,
\quad
(0.87)^{55}=0.0005.
\end{equation}

\vspace{3mm}\noindent
{\bf Problem}\ \ Clarify a relation between $I_{n}$ and $(0.87)^{n}$.

\section{Counter Example}
We show an example which does not give the PageRank 
vector in terms of the power method with usual initial vector 
${\bf e}_{1}$ when $H$ is not a realistic Google matrix.

\noindent
Example : a collection of web pages with links

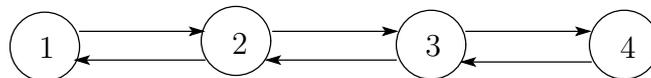
\vspace{5mm}
\begin{center}
\input{google-figure-2.tex}
\end{center}

\vspace{3mm}\noindent
The Google matrix for this graph is given by
\begin{equation}
H=
\left(
\begin{array}{cccc}
0 & \frac{1}{2} & 0 & 0 \\
1 & 0 & \frac{1}{2} & 0 \\
0 & \frac{1}{2} & 0 & 1 \\
0 & 0 & \frac{1}{2} & 0 
\end{array}
\right).
\end{equation}

The characteristic polynomial of $H$ is given by
\begin{equation}
f(\lambda)
=|\lambda E-H|
=\lambda^{4}-\frac{5}{4}\lambda^{2}+\frac{1}{4}
=(\lambda^{2}-1)(\lambda^{2}-\frac{1}{4})
\end{equation}
and the solutions are
\begin{equation}
\lambda=\pm 1,\ \pm \frac{1}{2}.
\end{equation}
Therefore, $H$ is not a realistic Google matrix because of 
$\lambda=-1$. See (\ref{eq:realistic}) once more. 
For $H$ it is easy to see that the PageRank vector is given by
\begin{equation}
I=
\left(
\begin{array}{c}
\frac{1}{6} \\
\frac{2}{6} \\
\frac{2}{6} \\
\frac{1}{6}
\end{array}
\right)
\equiv 
\left(
\begin{array}{c}
0.1667 \\
0.3333 \\
0.3333 \\
0.1667
\end{array}
\right).
\end{equation}

We show that $I$ is not obtained by the power method. 
In fact, it is easy to see
\begin{equation}
I_{2n}=H^{2n}{\bf e}_{1}=
\left(
\begin{array}{c}
a_{n} \\
0 \\
c_{n} \\
0
\end{array}
\right)
\quad \mbox{and}\quad 
I_{2n+1}=H^{2n+1}{\bf e}_{1}=
\left(
\begin{array}{c}
0 \\
b_{n} \\
0 \\
d_{n}
\end{array}
\right)
\end{equation}
where we don't need exact values of $a_{n},b_{n},c_{n},d_{n}$. 
As a result, $\{I_{n}\}$ does not converge. 

Next, as a trial we change the initial vector.  For example 
we set 
\begin{equation}
J_{n}=H^{n}J_{0}\quad \mbox{and}\quad 
J_{0}=
\left(
\begin{array}{c}
\frac{1}{4} \\
\frac{1}{4} \\
\frac{1}{4} \\
\frac{1}{4}
\end{array}
\right)
\end{equation}
because of $H^{T}J_{0}=J_{0}$. Let us list the calculation :
\begin{equation}
J_{10}=
\left(
\begin{array}{c}
0.1667 \\
0.3333 \\
0.3333 \\
0.1667
\end{array}
\right), \quad
J_{11}=
\left(
\begin{array}{c}
0.1666 \\
0.3334 \\
0.3334 \\
0.1666
\end{array}
\right), \quad
J_{12}=
\left(
\begin{array}{c}
0.1667 \\
0.3333 \\
0.3333 \\
0.1667
\end{array}
\right).
\end{equation}
From the result $n=10$ is enough.

Last, we present an important

\vspace{3mm}\noindent
{\bf Problem}\ \ We speculate that $J_{0}=(1/n,1/n,\cdots,1/n)^{T}$ 
is in general better than ${\bf e}_{1}=(1,0,\cdots,0)^{T}$ as an initial 
vector. Study this point in detail.

\vspace{10mm}\noindent 
{\bf Acknowledgments}\\
We would like to thank Yasushi Homma and Ryu Sasaki 
for useful suggestions and comments.


\end{document}

%% file: google-figure-1.tex
{\unitlength 0.1in%
\begin{picture}( 34.1000, 16.4900)( 28.0000,-21.9900)%
%
\special{pn 8}%
\special{ar 5010 1009 200 200  0.0000000  6.2831853}%
%
\special{pn 8}%
\special{ar 6010 999 200 200  0.0000000  6.2831853}%
%
\special{pn 8}%
\special{ar 4010 1000 200 200  0.0000000  6.2831853}%
%
\special{pn 8}%
\special{ar 3010 1000 200 200  0.0000000  6.2831853}%
%
\special{pn 8}%
\special{ar 5000 1999 200 200  0.0000000  6.2831853}%
%
\special{pn 8}%
\special{ar 6000 1989 200 200  0.0000000  6.2831853}%
%
\special{pn 8}%
\special{ar 4000 1990 200 200  0.0000000  6.2831853}%
%
\special{pn 8}%
\special{ar 3000 1990 200 200  0.0000000  6.2831853}%
%
\special{pn 8}%
\special{pa 3220 1000}%
\special{pa 3810 1000}%
\special{fp}%
\special{sh 1}%
\special{pa 3810 1000}%
\special{pa 3743 980}%
\special{pa 3757 1000}%
\special{pa 3743 1020}%
\special{pa 3810 1000}%
\special{fp}%
%
\special{pn 8}%
\special{pa 4220 1000}%
\special{pa 4810 1000}%
\special{fp}%
\special{sh 1}%
\special{pa 4810 1000}%
\special{pa 4743 980}%
\special{pa 4757 1000}%
\special{pa 4743 1020}%
\special{pa 4810 1000}%
\special{fp}%
%
\special{pn 8}%
\special{pa 4210 2000}%
\special{pa 4800 2000}%
\special{fp}%
\special{sh 1}%
\special{pa 4800 2000}%
\special{pa 4733 1980}%
\special{pa 4747 2000}%
\special{pa 4733 2020}%
\special{pa 4800 2000}%
\special{fp}%
%
\special{pn 8}%
\special{pa 3010 1200}%
\special{pa 3010 1780}%
\special{fp}%
\special{sh 1}%
\special{pa 3010 1780}%
\special{pa 3030 1713}%
\special{pa 3010 1727}%
\special{pa 2990 1713}%
\special{pa 3010 1780}%
\special{fp}%
%
\special{pn 8}%
\special{pa 5010 1210}%
\special{pa 5010 1790}%
\special{fp}%
\special{sh 1}%
\special{pa 5010 1790}%
\special{pa 5030 1723}%
\special{pa 5010 1737}%
\special{pa 4990 1723}%
\special{pa 5010 1790}%
\special{fp}%
%
\special{pn 8}%
\special{pa 3850 1120}%
\special{pa 3100 1810}%
\special{fp}%
\special{sh 1}%
\special{pa 3100 1810}%
\special{pa 3163 1780}%
\special{pa 3139 1774}%
\special{pa 3136 1750}%
\special{pa 3100 1810}%
\special{fp}%
%
\special{pn 8}%
\special{pa 4110 1810}%
\special{pa 4860 1130}%
\special{fp}%
\special{sh 1}%
\special{pa 4860 1130}%
\special{pa 4797 1160}%
\special{pa 4820 1166}%
\special{pa 4824 1190}%
\special{pa 4860 1130}%
\special{fp}%
%
\special{pn 8}%
\special{pa 5200 930}%
\special{pa 5820 930}%
\special{fp}%
\special{sh 1}%
\special{pa 5820 930}%
\special{pa 5753 910}%
\special{pa 5767 930}%
\special{pa 5753 950}%
\special{pa 5820 930}%
\special{fp}%
%
\special{pn 8}%
\special{pa 5830 1090}%
\special{pa 5200 1090}%
\special{fp}%
\special{sh 1}%
\special{pa 5200 1090}%
\special{pa 5267 1110}%
\special{pa 5253 1090}%
\special{pa 5267 1070}%
\special{pa 5200 1090}%
\special{fp}%
%
\special{pn 8}%
\special{pa 3200 1927}%
\special{pa 3820 1927}%
\special{fp}%
\special{sh 1}%
\special{pa 3820 1927}%
\special{pa 3753 1907}%
\special{pa 3767 1927}%
\special{pa 3753 1947}%
\special{pa 3820 1927}%
\special{fp}%
%
\special{pn 8}%
\special{pa 3810 2080}%
\special{pa 3180 2080}%
\special{fp}%
\special{sh 1}%
\special{pa 3180 2080}%
\special{pa 3247 2100}%
\special{pa 3233 2080}%
\special{pa 3247 2060}%
\special{pa 3180 2080}%
\special{fp}%
%
\special{pn 8}%
\special{pa 5200 1907}%
\special{pa 5820 1907}%
\special{fp}%
\special{sh 1}%
\special{pa 5820 1907}%
\special{pa 5753 1887}%
\special{pa 5767 1907}%
\special{pa 5753 1927}%
\special{pa 5820 1907}%
\special{fp}%
%
\special{pn 8}%
\special{pa 5830 2067}%
\special{pa 5200 2067}%
\special{fp}%
\special{sh 1}%
\special{pa 5200 2067}%
\special{pa 5267 2087}%
\special{pa 5253 2067}%
\special{pa 5267 2047}%
\special{pa 5200 2067}%
\special{fp}%
%
\special{pn 8}%
\special{pa 6100 1180}%
\special{pa 6100 1810}%
\special{fp}%
\special{sh 1}%
\special{pa 6100 1810}%
\special{pa 6120 1743}%
\special{pa 6100 1757}%
\special{pa 6080 1743}%
\special{pa 6100 1810}%
\special{fp}%
%
\special{pn 8}%
\special{pa 5990 800}%
\special{pa 5990 550}%
\special{fp}%
%
\special{pn 8}%
\special{pa 5990 550}%
\special{pa 3000 550}%
\special{fp}%
%
\special{pn 8}%
\special{pa 3000 550}%
\special{pa 3000 780}%
\special{fp}%
\special{sh 1}%
\special{pa 3000 780}%
\special{pa 3020 713}%
\special{pa 3000 727}%
\special{pa 2980 713}%
\special{pa 3000 780}%
\special{fp}%
\put(29.9000,-10.1000){\makebox(0,0){1}}%
\put(29.7000,-20.4000){\makebox(0,0)[lb]{2}}%
\put(39.6000,-10.5000){\makebox(0,0)[lb]{3}}%
\put(39.8000,-20.5000){\makebox(0,0)[lb]{4}}%
\put(49.8000,-10.6000){\makebox(0,0)[lb]{5}}%
\put(59.8000,-10.6000){\makebox(0,0)[lb]{7}}%
\put(49.8000,-20.6000){\makebox(0,0)[lb]{6}}%
\put(59.7000,-20.4000){\makebox(0,0)[lb]{8}}%
%
\special{pn 8}%
\special{pa 5150 1170}%
\special{pa 5840 1850}%
\special{fp}%
\special{sh 1}%
\special{pa 5840 1850}%
\special{pa 5807 1789}%
\special{pa 5802 1813}%
\special{pa 5778 1817}%
\special{pa 5840 1850}%
\special{fp}%
%
\special{pn 8}%
\special{pa 5920 1780}%
\special{pa 5920 1170}%
\special{fp}%
\special{sh 1}%
\special{pa 5920 1170}%
\special{pa 5900 1237}%
\special{pa 5920 1223}%
\special{pa 5940 1237}%
\special{pa 5920 1170}%
\special{fp}%
\end{picture}}%

%% file: google-figure-2.tex
{\unitlength 0.1in%
\begin{picture}( 33.6000,  3.9000)( 26.2000,-11.8000)%
%
\special{pn 8}%
\special{ar 2800 1000 180 180  0.0000000  6.2831853}%
%
\special{pn 8}%
\special{ar 3790 970 180 180  0.0000000  6.2831853}%
%
\special{pn 8}%
\special{ar 4800 990 180 180  0.0000000  6.2831853}%
%
\special{pn 8}%
\special{pa 2980 920}%
\special{pa 3620 920}%
\special{fp}%
\special{sh 1}%
\special{pa 3620 920}%
\special{pa 3553 900}%
\special{pa 3567 920}%
\special{pa 3553 940}%
\special{pa 3620 920}%
\special{fp}%
%
\special{pn 8}%
\special{pa 3980 920}%
\special{pa 4620 920}%
\special{fp}%
\special{sh 1}%
\special{pa 4620 920}%
\special{pa 4553 900}%
\special{pa 4567 920}%
\special{pa 4553 940}%
\special{pa 4620 920}%
\special{fp}%
%
\special{pn 8}%
\special{pa 4980 920}%
\special{pa 5620 920}%
\special{fp}%
\special{sh 1}%
\special{pa 5620 920}%
\special{pa 5553 900}%
\special{pa 5567 920}%
\special{pa 5553 940}%
\special{pa 5620 920}%
\special{fp}%
%
\special{pn 8}%
\special{pa 5630 1080}%
\special{pa 4960 1080}%
\special{fp}%
\special{sh 1}%
\special{pa 4960 1080}%
\special{pa 5027 1100}%
\special{pa 5013 1080}%
\special{pa 5027 1060}%
\special{pa 4960 1080}%
\special{fp}%
%
\special{pn 8}%
\special{pa 4620 1070}%
\special{pa 3950 1070}%
\special{fp}%
\special{sh 1}%
\special{pa 3950 1070}%
\special{pa 4017 1090}%
\special{pa 4003 1070}%
\special{pa 4017 1050}%
\special{pa 3950 1070}%
\special{fp}%
%
\special{pn 8}%
\special{pa 3630 1070}%
\special{pa 2960 1070}%
\special{fp}%
\special{sh 1}%
\special{pa 2960 1070}%
\special{pa 3027 1090}%
\special{pa 3013 1070}%
\special{pa 3027 1050}%
\special{pa 2960 1070}%
\special{fp}%
\put(27.7000,-10.6000){\makebox(0,0)[lb]{1}}%
\put(37.7000,-10.4000){\makebox(0,0)[lb]{2}}%
\put(47.7000,-10.5000){\makebox(0,0)[lb]{3}}%
%
\special{pn 8}%
\special{ar 5800 990 180 180  0.0000000  6.2831853}%
\put(57.8000,-10.6000){\makebox(0,0)[lb]{4}}%
\end{picture}}%

%% file: Google_Matrices.bbl
\begin{thebibliography}{99}
%
%
\bibitem{DA}David Austin : 
\newblock How Google Finds Your Needle in the Web's Haystack, 
\newblock Feature Column, monthly essays on mathematical topics, 
\newblock http://www.ams.org/samplings/feature-column/fcarc-pagerank.
%
\bibitem{SG}Semyon Gerschgorin : 
\newblock {\"U}ber die Abgrenzung der Eigenwerte einer Matrix, 
\newblock Izv. Akad. Nauk. USSR Otd. Fiz.-Mat. Nauk 6 (1931), 749-754. 
\newblock Honestly writing, we have not seen this paper.
%
\bibitem{IS}Ichiro Satake : 
\newblock Linear Algebra (in Japanese), 
\newblock Shokabo, Tokyo, 1975.
%
\bibitem{KF}Kazuyuki Fujii : 
\newblock Introduction to Linear Algebra (in Japanese), 
\newblock Lecture note at Yokohama City University, 2014.
%
%
\end{thebibliography}
